# Photoresponse properties of single-crystalline thick film based on high-entropy topological insulator $(Bi_{3/4}Sb_{1/4})_2(Te_{2/5}Se_{2/5}S_{1/5})_3$

Alexei Vasil'ev[a], Marina Zhezhu[a], Oleg Ivanov[b]

[a]A.B. Nalbandyan Institute of Chemical Physics NAS RA, Yerevan 0014, Armenia

[b]Belgorod State University, Belgorod 394015, Russia

Corresponding author: E-mail: vasilev.alexei@ichph.sci.am

## Abstract

High-entropy topological insulator $(Bi_{3/4}Sb_{1/4})_2(Te_{2/5}Se_{2/5}S_{1/5})_3$ has been for the first time prepared by self-propagating high-temperature synthesis and melting methods. Single-crystalline thick-filmed sample with thickness of ~0.1 mm was applied to examine the photoresponse properties by using room-temperature chopped-light technique. Photodetector based on $(Bi_{3/4}Sb_{1/4})_2(Te_{2/5}Se_{2/5}S_{1/5})_3$ demonstrated strong photovoltaic response under irradiation of a white LED light, covering 420-730 nm range. The photocurrent increases from ~20 to ~55 μA with the increase in intensity from 100 to 600 W/m$^2$ that is originated from the increase in the number of non-equilibrium carriers generated by light. Room-temperature responsivity and detectivity of the photodetector are ~8.7 mA/W and ~10$^9$ Jones, respectively.

## 1. Introduction

Currently, topological insulators (TIs) are attractive and promising materials for both fundamental condensed matter physics and for various applications in high-speed dissipationless electronics, spintronics and quantum computing [1,2]. Key features in the TIs properties are due to combination and interaction of bulk insulating band gap states and surface conducting gapless states with a linear Dirac dispersion. Owing to linear dispersion relationship, high carrier mobility, and special energy band structure, TIs are also considered as promising materials for the photodetectors [3]. One of famous TIs is bismuth telluride, $Bi_2Te_3$ [4]. $Bi_2Te_3$ is often applied as a parent compound to synthesize high-entropy alloys (HEA) [5]. The HEAs consist of five or more principal elements taken in equimolar or near-equimolar ratios, competing for the same position in crystal lattice. Due to effects of the high-mixing entropy and chemical complexity, the HEAs demonstrate many unique properties, useful for variety of applications [6]. Today, developing HEAs is effective approach that allows improving the properties of various materials.

Five-component $(Bi_{3/4}Sb_{1/4})_2(Te_{2/5}Se_{2/5}S_{1/5})_3$ compound, which was also derived from the parent $Bi_2Te_3$, is new HEA. According to our recent study, thick-filmed $(Bi_{2/3}Sb_{1/3})_2(Te_{2/5}Se_{2/5}S_{1/5})_3$ sample demonstrated features in the properties, specific for TIs [7]. Since $(Bi_{3/4}Sb_{1/4})_2(Te_{2/5}Se_{2/5}S_{1/5})_3$ and $(Bi_{2/3}Sb_{1/3})_2(Te_{2/5}Se_{2/5}S_{1/5})_3$ are only slightly different in a Bi/Sb ratio, $(Bi_{3/4}Sb_{1/4})_2(Te_{2/5}Se_{2/5}S_{1/5})_3$ can be reasonably assumed to be TI. Hence, $(Bi_{3/4}Sb_{1/4})_2(Te_{2/5}Se_{2/5}S_{1/5})_3$ should be tested as promising photodetector material. The aim of this work is to examine the photoresponse properties of high-entropy topological insulator $(Bi_{3/4}Sb_{1/4})_2(Te_{2/5}Se_{2/5}S_{1/5})_3$, prepared as single-crystalline thick-filmed sample.

## 2. Materials and Methods

Self-propagating high-temperature synthesis (SHS) was applied to synthesize a starting $(Bi_{3/4}Sb_{1/4})_2(Te_{2/5}Se_{2/5}S_{1/5})_3$ compound. To prepare the reaction powder mixture for SHS process, desired amounts of Bi, Sb, Se, Te and S powders were thoroughly mixed in a mortar for 1 h in a hexane acting as a liquid media. To prevent deviations from the desired $(Bi_{3/4}Sb_{1/4})_2(Te_{2/5}Se_{2/5}S_{1/5})_3$ composition due to high-temperature evaporation, chalcogens were taken with 5% molar excess. The reaction mixture was dried from residual solvent under vacuum at room temperature for 1 h. Then, the mixture was compacted into cylinders of 20 mm in diameter and 20 mm in height and placed in quartz tube. The SHS process was initiated by rapidly igniting the cylinder with handheld torch. After the process started, the torch was removed allowing the SHS reaction to spread throughout the entire reaction volume. The SHS process was carried out under a constant vacuum of ~0.02 Torr using a fore vacuum pump. After natural cooling, the SHS-synthesized material was thoroughly ground in the mortar for 1 h. The resulting powder was put in quartz tube, which was under continuous evacuating by a pump. Under heating, the powder was totally melted at 1273 K. After holding at this temperature for 10 min, the melt was naturally cooled to room temperature, resulting in ingot. The ingot consisted of many individual and randomly oriented mm-sized crystals, fused with each other(Fig. S1). Using adhesive tape, filmed samples could be extracted from a surface of one of individual crystals. The thick-filmed sample with 4.0×2.0×0.1 mm sizes was applied for study of photorsponse properties. Since length of the film (4.0 mm) is far bigger than its thickness (0.1 mm), the sample can be considered as 2D-system. Applying 2D-films based on TIs materials is necessary to enhance surface-to-volume ratio that will also enhance the surface states contribution into the properties of TIs.

To find phase composition and confirim single-crystallinity of the sample, X-ray diffraction (XRD) examination was carried out by using a Rigaku Miniflex diffractometer with $CuK_{\alpha}$-radiation. Scanning electron microscopy (SEM, Prisma microscope) was applied for characterizing an elemental composition of the filmed sample being studied by surface mapping of various elements distributions (energy dispersive X-ray spectroscopy, SEM-EDS method). Raman spectrometry (Horiba LabRAM Nano AFM-Raman System) allowed investigating features in the phonon modes due to specific defect structure of the sample. The Ecopia HMS-5000 was used to determine the transport properties (Table S1). To extract photoresponse properties of the sample, a Zahner CIMPS-QE/IPCE-UV Photovoltaic/Photoelectrochemical System was involved. A light source is white LED, emitting 420-730 nm light. To study the photoresponse properties, a simple photodetector device based on the thick-filmed $(Bi_{3/4}Sb_{1/4})_2(Te_{2/5}Se_{2/5}S_{1/5})_3$ sample with two electrodes attached by conductive glue was created. Linear I-V curve of the device confirmed formation of good ohmic contact. Chopped-light technique was applied at room temperature under varying light intensities and zero bias voltage.

**3. Results and Discussion**

Part of the ingot that was applied to extract the filmed sample was ground to prepare powder material. XRD pattern for this material is shown in Fig. 1 (a).

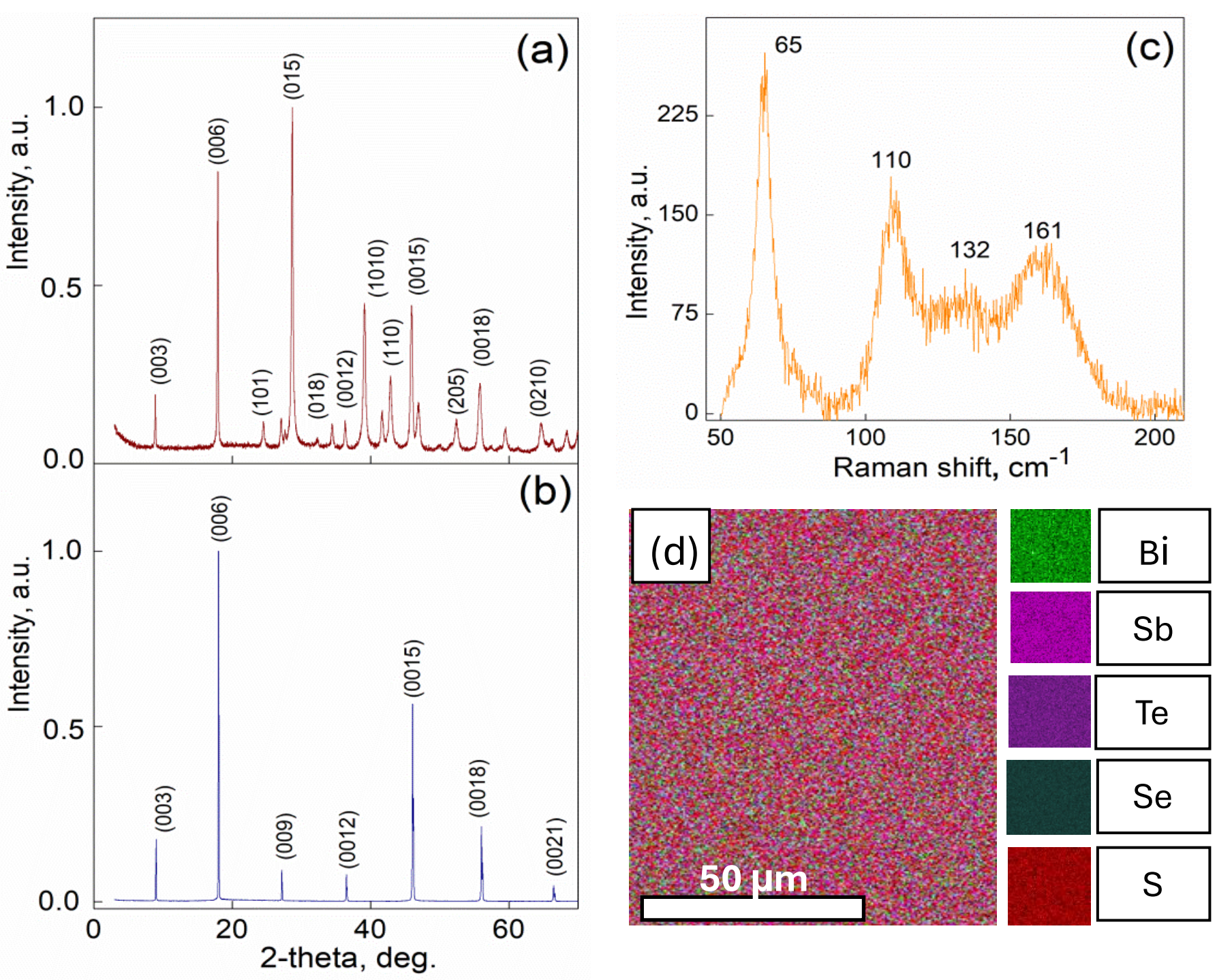

**Fig. 1** XRD patterns taken for powder $(Bi_{3/4}Sb_{1/4})_2(Te_{2/5}Se_{2/5}S_{1/5})_3$ material (a) and from (*a*-*b*) plane surface of the filmed $(Bi_{3/4}Sb_{1/4})_2(Te_{2/5}Se_{2/5}S_{1/5})_3$ sample (b), Raman spectrum of powder material (c), and SEM-EDS map of different elements for (*a*-*b*) plane surface (d)

The pattern corresponds to single hexagonal phase with crystal $R\bar{3}m$ structure and lattice $a$=$b$=4.212 Å and $c$=29.609 Å parameters. This structure is typical for $Bi_2Te_3$ [7]. XRD pattern taken from larger 4.0×2.0 mm surface of the filmed sample is shown in Fig. 1 (b). Only (00*l*) peaks, originated from (*a*-*b*) plane reflection, are observed in XRD pattern. Therefore, the filmed sample is single-crystalline, and larger surface of the film is oriented parallel to crystal (*a*-*b*) plane. Formation of $R\bar{3}m$ structure in the sample was also confirmed by Raman spectroscopic analysis. Raman spectrum is typical for the $R\bar{3}m$ structure (Fig. 1 (d)). Two peaks at 65 and 161 $cm^{-1}$ are related to perpendicular $A_{1g}$ and $A_{2g}$ vibrational modes, while oscillation at 110 $cm^{-1}$ is related to planar $E_g$ mode. Vibrational mode, detected at 132 $cm^{-1}$, was earlier observed in $Bi_2SeTe_2$ [8]. This mode was attributed to localized defect-induced $V_1$ mode due to forming Te-Se anti-site point defects. According to SEM-EDS mapping results, Bi, Sb, Te, Se and S are homogeneously distributed on surface of the sample (Fig. 1 (c)). Elemental composition was determined as 30 at.% Bi, 11 at.% Sb, 25 at.% Te, 24 at.% Se and 10 at.% S that well corresponds to the desired $(Bi_{3/4}Sb_{1/4})_2(Te_{2/5}Se_{2/5}S_{1/5})_3$ composition(Fig. S2).

Photodetector based on the $(Bi_{3/4}Sb_{1/4})_2(Te_{2/5}Se_{2/5}S_{1/5})_3$ sample demonstrated a remarkable photoresponse, strongly depending on light illumination intensity, $P$ (Fig. 2 (a)).

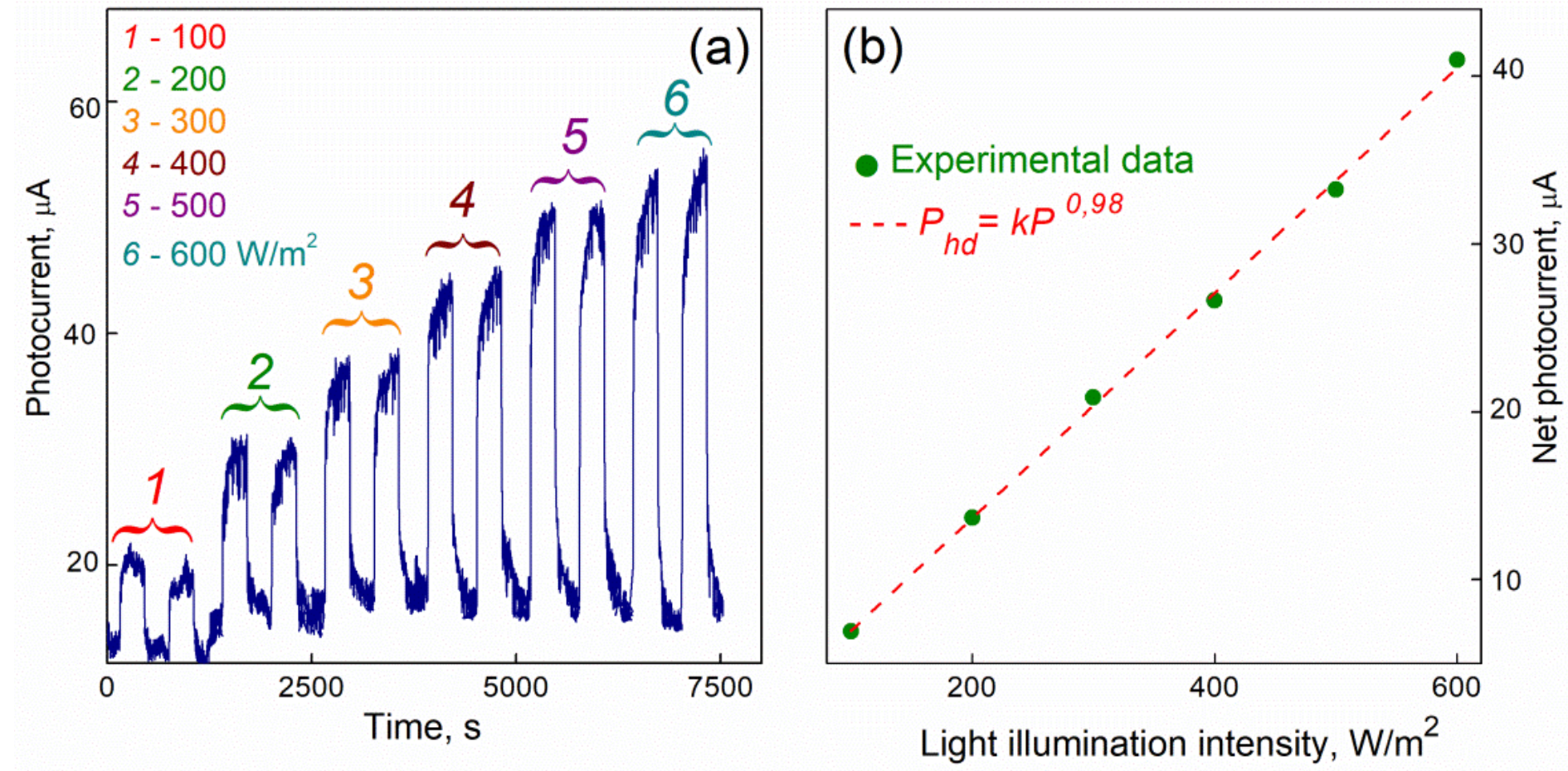

**Fig.2** (a) Photoresponse of photodetector based on the $(Bi_{3/4}Sb_{1/4})_2(Te_{2/5}Se_{2/5}S_{1/5})_3$ sample to the on/off white LED irradiation, (b) Dependence of net photocurrent on the light illumination intensity

Under light irradiation conditions, the photocurrent, $I_p$, rises immediately, and drops sharply when it is sheltered. The photocurrent expectedly increases from ~20 to ~55 μA with the increase in $P$ from 100 to 600 $W \cdot m^{-2}$ that is originated from the increase in the number of non-equilibrium carriers generated by light. To confirm the reproducibility of results, the photocurrent was measured twice at the same intensity. The results for both measurements were very close to each other. Data of Fig. 2 (a) were applied to extract the net photocurrent ($I_{ph}=I_{illuminated}$ - $I_{dark}$, where $I_{illuminated}$ and $I_{dark}$ are the currents obtained under illumination and dark conditions, respectively). The $I_{ph}(P)$ dependence for photodetectors are described by expression $I_{ph} = kP^{\theta}$, where $k$ and $\theta$ are parameters, depending on the illumination wavelength and photocurrent generation mechanism, respectively. Ideal photodetector operating under a photovoltaic mechanism corresponds to $\theta$=1. As shown by dashed line, the $I_{ph}(P)$ dependence for $(Bi_{3/4}Sb_{1/4})_2(Te_{2/5}Se_{2/5}S_{1/5})_3$ sample can be fitted with $\theta\approx0.98$, i.e. the photodetector based on high-entropy topological insulator $(Bi_{3/4}Sb_{1/4})_2(Te_{2/5}Se_{2/5}S_{1/5})_3$ behaves very close to ideal photodetector. Then, to characterize the photodetector, rise time and decay time (see the Supplementary Information), responsivity, $R$, and detectivity, $D$, were calculated. Responsivity, calculated by $R = I_{ph}/PA$, characterizes of a photodetector's response to incident light. Detectivity, calculated by $D = R \times \sqrt{S/2eI_{dark}}$ , characterizes a photodetector's ability to sense weak signals in the detector noise. The $R(P)$ and $D(P)$ dependences are shown in Fig. 3 (*1*) and (*2*) curves, respectively.

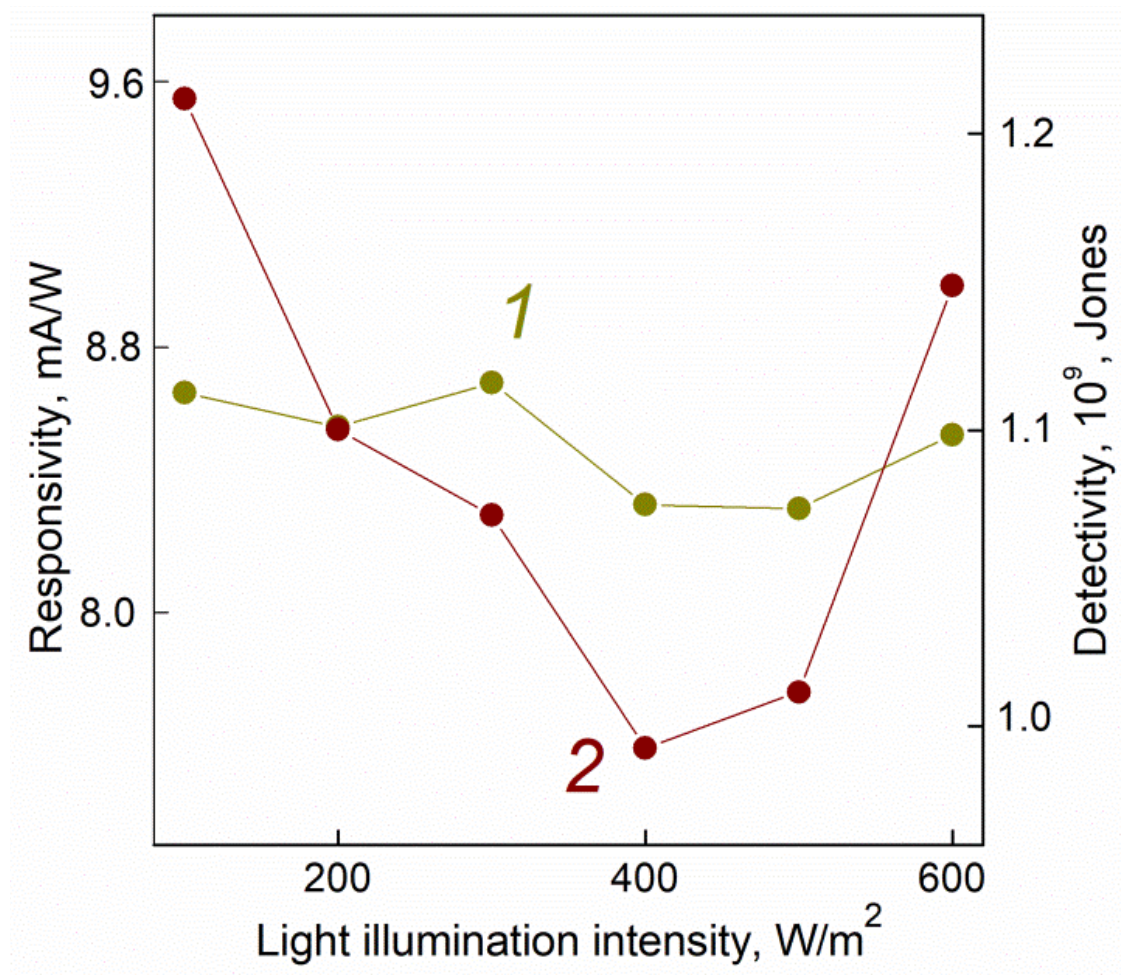

**Fig. 3** Calculated responsivity (curve *1*) and detectivity (*2*) under various light intensities

The responsivity is weakly $P$-dependent and randomly scatters around average value $R\approx8.7$ mA/W. This value is in accordance with the $R$ values of some photodetectors based on TI materials, listed in Refs. [3, 9] ($R$=2.2 mA/W for $Bi_2Te_3$/Graphene, 1 mA/W for $Bi_2Te_{2.7}Se_{0.3}$/Si nanoplates, 2.45 mA/W for pristine $Bi_2Se_3$ bulk, 16.1 mA/W for heat-treated $Bi_2Se_3$ nanosheets, 20.4 mA/W for $Bi_2Se_3$ nanosheets (exfoliated), etc.). The detectivity of our photodetector is ~$10^9$ Jones. This value falls in the $10^8$-$10^{13}$ Jones range, which corresponds to the detectivities of many the photodetector's materials [1]. The $D$ is firstly falling up to $P$=400 W/m$^2$, and then starts increasing. Since $Bi_2Te_3$-based HEAs are effective thermoelectrics [10], the minimum in the $D(P)$ dependence can be related to a thermoelectric response, originated from gradient heating of the photodetector's material during the light irradiation.

**Conclusion**

Thus, single-crystalline thick-filmed $(Bi_{3/4}Sb_{1/4})_2(Te_{2/5}Se_{2/5}S_{1/5})_3$ material with homogeneous Bi, Sb, Te, Se and S distribution, combining properties of high-entropy alloys and topological insulators, has been for the first time prepared. The photoresponse properties of the $(Bi_{3/4}Sb_{1/4})_2(Te_{2/5}Se_{2/5}S_{1/5})_3$-based photodetector was tested by using room-temperature chopped-light technique under irradiation of white LED light, covering 420-730 nm range. Room-temperature responsivity and detectivity of the photodetector were found to be ~8.7 mA/W and ~$10^9$ Jones, respectively. Taking into account these photoresponse properties, high-entropy topological insulator $(Bi_{3/4}Sb_{1/4})_2(Te_{2/5}Se_{2/5}S_{1/5})_3$ can be considered as promising material for photovoltaic applications.

**Supplementary Information for**

# Photoresponse properties of single-crystalline thick film based on high-entropy topological insulator $(Bi_{3/4}Sb_{1/4})_2(Te_{2/5}Se_{2/5}S_{1/5})_3$

Alexei Vasil'ev[a], Marina Zhezhu[a], Oleg Ivanov[b]
[a]A.B. Nalbandyan Institute of Chemical Physics NAS RA, Yerevan 0014, Armenia
[b]Belgorod State University, Belgorod 394015, Russia
Corresponding author: E-mail: vasilev.alexei@ichph.sci.am

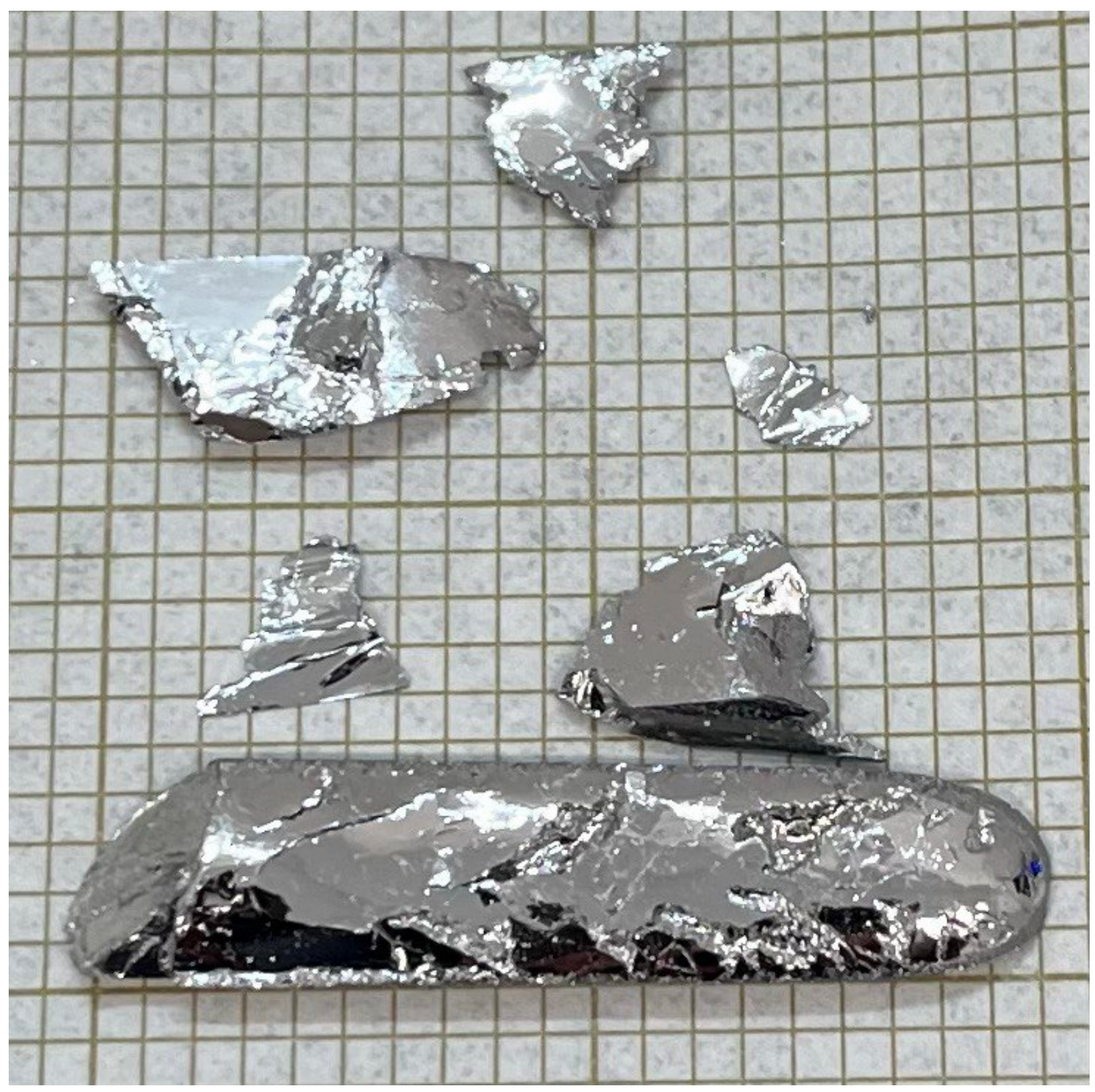

Fig. S1 Photo of the single-crystalline $(Bi_{3/4}Sb_{1/4})_2(Te_{2/5}Se_{2/5}S_{1/5})_3$ ingot

The synthesized crystal has been found to be stable under ambient air conditions.

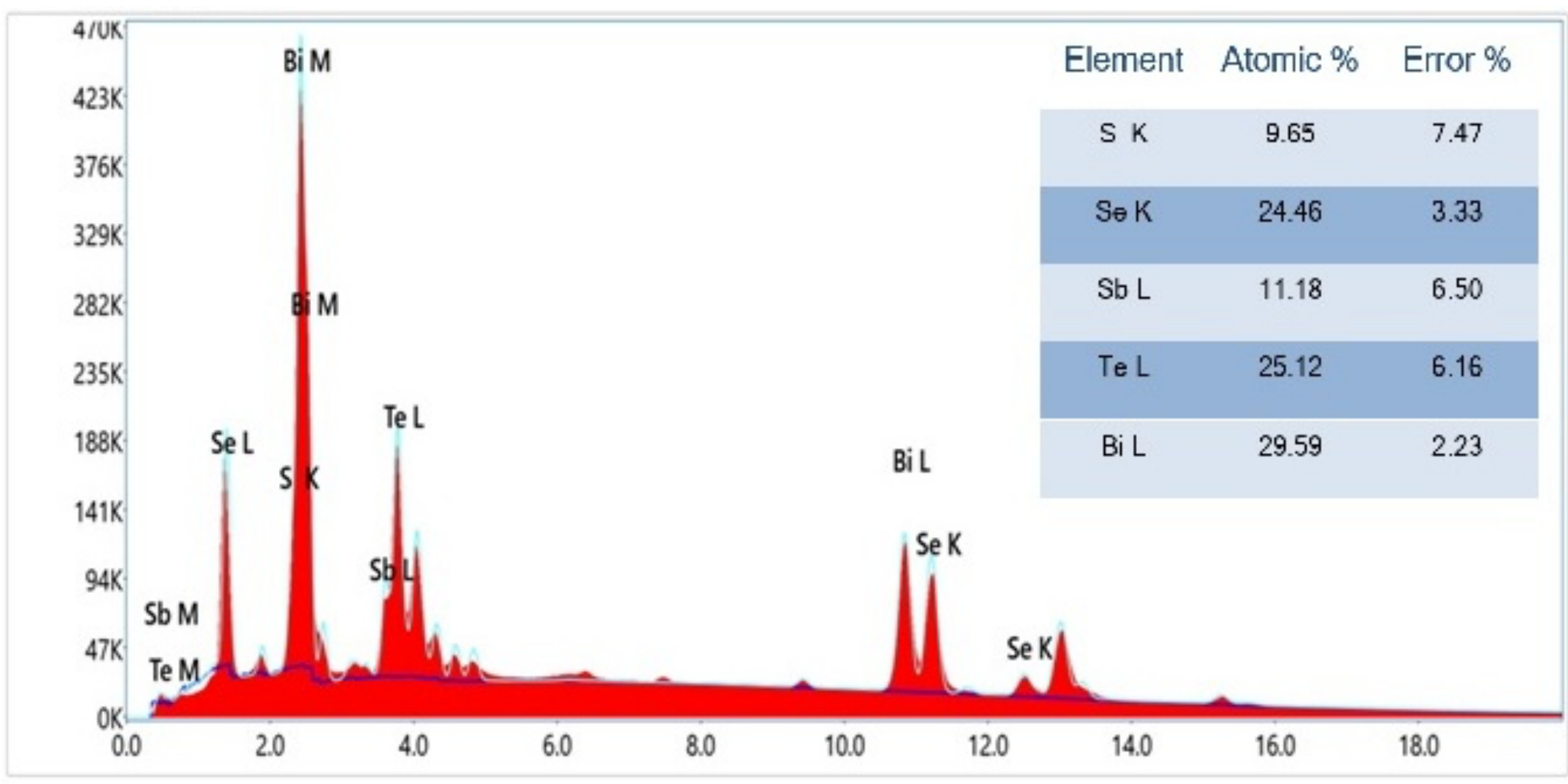

Fig. S2 EDS analysis of the $(Bi_{3/4}Sb_{1/4})_2(Te_{2/5}Se_{2/5}S_{1/5})_3$

*Details of the photoresponse time for the thick-film sample*

To evaluate the response speed of the device (defined as the time required for the photocurrent to increase from 10% to 90% of its peak value, or decrease from 90% to 10%, referred to as the rise time and decay time, respectively), the photoresponse data were analyzed and averaged. The rise and decay times were estimated to be 100 s and 75 s, respectively.

Compared to the literature data, these are relatively long rise and decay times, which may indicate the presence of trap centers formed as a result of antisite (Fig. 1(c)) and other defects in the single crystal, potentially reducing the response speed.

Table S1. Transport properties of the thick-filmed $(Bi_{3/4}Sb_{1/4})_2(Te_{2/5}Se_{2/5}S_{1/5})_3$ at room temperature

| Resistivity (μOhm/m) | Hall concentration ($cm^{-3}$) | Hall Mobility ($cm^2/V/s$) |
|---|---|---|
| 23.8 | $4.84*10^{18}$ | 542 |

*Details of the transport properties*

The Hall constant determined from the transport measurements was negative, indicating that electrons are the majority charge carriers. The carrier concentration was estimated to be $4.84*10^{18}$ $cm^{-3}$. The resistivity of the sample is comparable in magnitude to that of topological materials based on bismuth telluride, due to the characteristically narrow bandgap of such compounds. The relatively high charge carrier mobility positively influences the transport of photogenerated carriers, contributing significantly to the photoresponse. However, the calculated long photoresponse times indicate the presence of trap centers, which degrade the photodetector performance.